\documentclass{IEEEcsmag}

\usepackage[colorlinks,urlcolor=blue,linkcolor=blue,citecolor=blue]{hyperref}

\usepackage{upmath}
 \usepackage[frozencache,cachedir=.]{minted}
\jvol{XX}
\jnum{XX}
\paper{8}
\jmonth{May/June}
\jname{Software}
\pubyear{2019}

\usepackage{natbib}
\begin{document}

\sptitle{Department: SE4AI}
\editor{Editor: Tim Menzies, timm@ieee.org}

\title{AI-driven Development Is Here: Should You Worry?}

\author{Neil A. Ernst}
\affil{University of Victoria}

\author{Gabriele Bavota}
\affil{Università della Svizzera italiana}


\markboth{Department Head}{\title}

\maketitle

{\em From the Editor:

The benefits, as well as the draw-backs,  of  new AI technology
needs to be   carefully scrutinized. For example, in this article,
researchers review the promise, and potential pitfalls,
of AI tools that help programmer co-write their code.

This ``SE for AI'' column publishes commentaries on the growing field of SE for AI. Submissions are welcomed and encouraged
(1,000–2,400 words, each figure and table counts as 250 words, try to use fewer
than 12 references, and keep the discussion practitioner focused). Please submit
your ideas to me at timm\@ieee.org.—Tim Menzies
}\newpage

\chapterinitial{AI-driven Development Environments (AIDEs)} integrate the power of modern AI into IDEs like Visual Studio Code and JetBrains IntelliJ. By leveraging massive language models and the plethora of openly available source code, AIDEs promise to automate many of the obvious, routine tasks in programming. At the same time, AIDEs come with new challenges to think about, such as bias, legal compliance, security vulnerabilities, and their impact on learn programming.

Time was, programmers were laboriously entering machine instructions on punch cards, to be painstakingly offered up to the machine for processing. Nowadays, of course, we type code into editors, often hosted online, and get near-instant feedback on the compilation/test outcomes of our latest change. Let's call this the Integrated Development Environment (IDE) Revolution. 

Another revolution is now underway, call it the ``AI-driven Development Environment" (AIDE), one where a programmer can leverage the hive-mind of the many many programmers that came before them. Large language models are now available to make suggestions based on what hundreds of previous programmers typed in.

Not only is the syntax of programming ``natural'', but in fact a lot of programming follows repetitive patterns. If you have used a form of autocomplete in your IDE, you have seen a very primitive version of this idea (autocomplete seems to have been popularized with Microsoft's Visual-* tools in the 90s). Autocomplete at its simplest looks for all valid completions, such as possible library calls in \texttt{List.}. But things have progressed very rapidly in this space. Think about preparing some data analysis code:

\begin{minted}[fontsize=\footnotesize]{python}
# import a file named "observations.csv" 
# and print the number of rows and columns
# then use the data to create 
# a plot of the observations
import csv
import matplotlib.pyplot as plt
import numpy as np

data = []

with open('observations.csv') as csvfile:
    reader = csv.reader(csvfile)
    for row in reader:
        data.append(float(row[0]))

print("Number of rows: ", len(data))
print("Number of columns: ", len(data[0]))

plt.plot(data)
plt.show()
\end{minted}

Spoiler alert: we didn't write a single line of that code! Instead, we typed the four lines of comments into Visual Studio Code, a lightweight IDE, and had GitHub's Copilot extension ``write" the actual Python. 

Copilot, and its underlying AI engine, OpenAI's Codex, leverages the power of large language models trained on billions of text data online. Codex further tunes those language models with publicly available GitHub repositories. In essence, since so many repositories contain similar lines to the ones we wrote above, the language model can associate those lines with the comments we wrote, and something that might seem like magic can appear. 

This did not happen overnight, of course. Codex is the product of: 
\begin{itemize}
	\item AI research into deep neural networks and language models;
	\item Software repository mining for extracting data from repositories;
	\item Investigations into the linguistic nature of source code \citep{naturalness};
	\item Massive amounts of compute and data storage.
\end{itemize}

Codex and Copilot are not alone: several other AIDEs exist, including TabNine\footnote{\href{https://www.tabnine.com/}{tabnine.com}} and Kite\footnote{\href{https://www.kite.com}{kite.com}}. Intelligent fill in spreadsheets is a close companion, as is program repair such as Facebook's SapFix~\citep{sapfix}.

Codex \citep{codex} is built on OpenAI's GPT-3 language model with 12B parameters, and fine-tuned on 159 Gb of data from public GitHub repositories. The nature of a language model is that it builds deep associations between words in a high-dimensional space, learning complex patterns that produce code. They do not require abstract representations of code like ASTs, working on the token level directly. With suitable unit test coverage, generating 100 possible completions produced a solve rate (pass@100) of over 70\% on a benchmark of common completion tasks.

Codex promises to dramatically change the way we write and build software systems. Like all AI technology, it has great promise, but also comes with several new challenges for research and for practice.

\section{Promises of AI-driven DEs}
{\bf Automate the mundane} Much of software development is routine. Developers get a bug report, track down the bug, and file a patch; they wire library code together to leverage APIs; they need to display database records on a web page and handle any updates. 
Much of software development is also staggeringly complex and creative, too! Software is, as Grady Booch once wrote, ``the invisible thread ... on which we weave the fabric of computing." A key developer task then is to carefully distinguish those tasks which are complex, and those which are obvious or complicated \citep{cynefin}. 
AIDEs can remove the accidental complexity from what are obvious tasks, just like the code showed earlier. An AIDE like Copilot is already capable of automating these routine tasks, and other technologies, such as the automated program repair work of Facebook's SapFix tool~\citep{sapfix}, are tackling similar routine tasks. 

{\bf Automate API interactions} Much of programming today is about framework and API-driven development: connecting to a third party service, processing the result, and sending the result back to the user. Just as often we work within an existing architectural framework, for example for web or mobile applications, and our programs are closely coupled with those library calls. Many of these library calls are routine and repetitive for each new variant of an app. 

{\bf Teach Programming} As programming languages and APIs proliferate, learning new approaches and syntax becomes more challenging. Stack Overflow is invaluable for specific answers to common, and not so common, programming problems. For example, how does one configure a particular plotting library such as R's \texttt{ggplot} to change the background colour? But Stack Overflow, while incredibly helpful (we certainly could not program without it), requires one to leave the IDE to ask questions or perform search. AIDE promises the knowledge potential of Stack Overflow while avoiding continuous context switches between the IDE and the browser. And this will be useful for novices and experts alike. No more `yak shaving'\footnote{Yak shaving refers to doing a series of trivial tasks which distract you from the original, and important, goal. Compare with \emph{bikeshedding}.} trying to figure out the correct series of syntax calls for a given problem.

\section{Possible Challenges with AIDE}
Like any software development, AIDE will come with a host of challenges to be overcome, challenges in traditional software concerns such as defects and security vulnerabilities, but in new areas as well.

{\bf Copyright and Licensing} Codex is trained on (54 million) public GitHub repositories, and the creators of these GitHub repositories agreed to Codex-like usage of their code. However, the Codex context of use was something most of us probably did not anticipate. Does Codex have the right to all the code it was trained on? For output the language model is a series of weights, so in theory, code produced is an amalgamation of the inputs. Accordingly to a recent study \citep{copilot-copying} Codex rarely quotes code verbatim from the training set and when this happens it is usually code largely reused across open source projects. Does Codex-created code violate copyright? Is it fair use? We don't have an answer to this question, and open source licenses might need to be revisited to explicitly regulate the usage of code for training commercial code generation tools. Also, currently Codex output is the intellectual work product of the person who activated Codex, but this is currently because Codex is a beta, and these terms might change.

{\bf Learning to Program} The nature of learning programming will change dramatically with AIDEs. Whether these assistants will speed up or slow down the learning process is currently an open question. On the one hand, novice programmers can benefit from AIDEs by receiving recommendations useful to deal with tasks they struggle with. On the other hand, the risk of not fully understanding the received recommendations and just accept them is there. On top of this, AIDEs do also pose challenges for instructors: Codex is already so good it might surpass first year university students in introduction to programming  (CS1) courses. Some initial results in our testing show it is relatively simple to get Codex to generate reasonable (passing) solutions. CS1 programming assignments must change to handle those students who can merely pass the entire assignment spec to Codex for a solution.

\begin{figure}
    \centering
    \includegraphics[width=\linewidth]{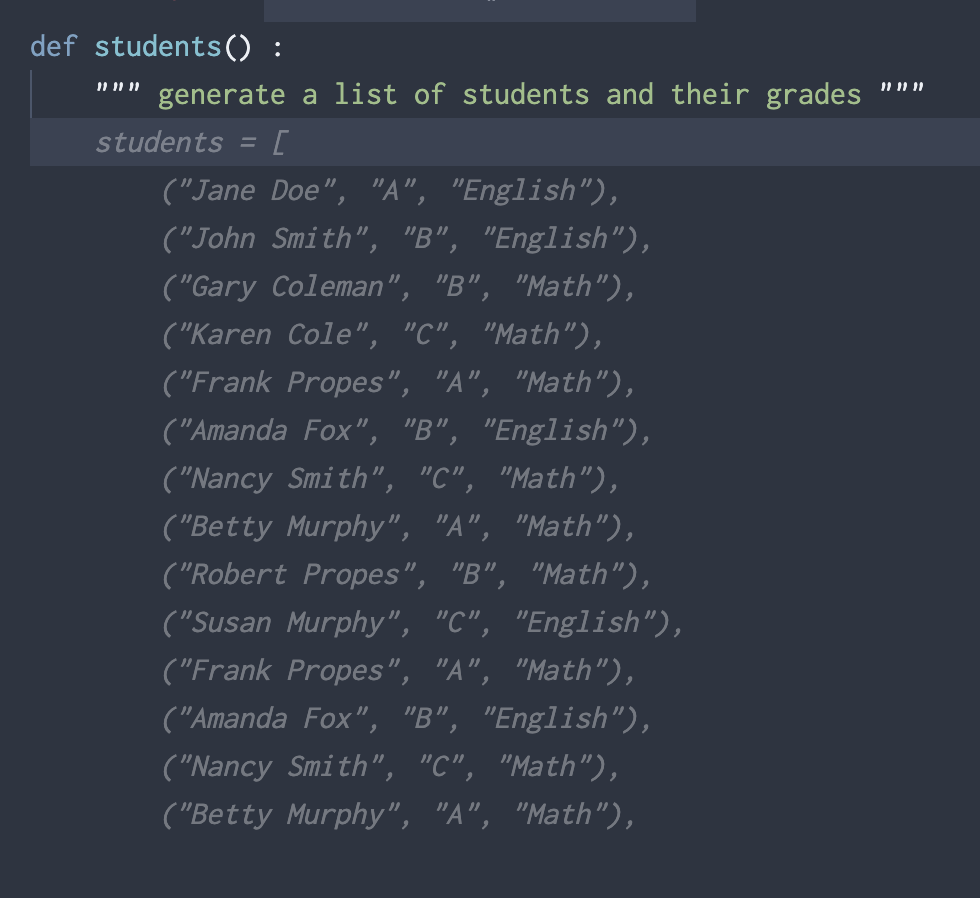}
    \caption{Copilot portends a new generation of AI-based productivity. The benefits, as well as the drawbacks, of this new approach need to be carefully scrutinized. 
    For example,
    asking Copilot  to  generate  a  list  of names   produces the gray text; which is a    list   of   predominantly   English/American  names (indicating an interesting, if not perhaps troubling,
    bias in its training data.)}
    \label{fig:names}
\end{figure}

{\bf Dataset Quality} Plenty of code freely available online has flaws. For example, much of it features student submissions, one-off explorations, or other low quality work \citep{pitfalls}. Like any trained model, Codex and other AIDEs are only as good as the training data. And although Codex has done extensive work filtering low-quality inputs, there remains code that has bugs, that has technical debt, or that uses outdated APIs. Subtle security holes can easily persist even in high-quality, high-volume repositories (consider the OpenSSH Heartbleed incident), and recent work showed how deep learning models can learn vulnerable code and inject it during autocompletion \citep{vulnerabilities}. AIDE demands that humans inspect its outputs carefully, but if we use it to create code for a problem we don't fully understand, we won't be able to understand its outputs either.  

More worrisome is that the language model reflects the biases that we humans have. For example, asking Copilot to generate a list of names produces a list of predominantly English/American names (Fig. \ref{fig:names}). Plotting suggestions generate graphs that fail to accommodate people with color-blindness. This is of course both a challenge for us, as much as it is for AIDE.

{\bf Sociotechnical Questions} The IDE revolution produced a now well known paradigm in computer programming, with continuous integration workflows dominant. But AIDE will possibly change that as more and more of the work is routinized and automated. More programmer time will be available for complex problem solving. But that means our current knowledge of how humans and machines interact will change. When Facebook rolled out their automated bug repair approach, one of the biggest challenges was not the technical problem, but rather integrating the repair bot into the humans that worked with them~\citep{scam_sapfix}.

{\bf Context and Complexity} Mechanization---such as in steel-making or automotive---has greatly improved productivity at the expense of those humans doing the routine. AIDEs will likely be no different. To what degree will an AIDE be able to carefully contextualize the solution for a specific problem? Where is the line between the routine and simple, and the complex and contextual? Will AI eventually design and write complex software solutions? Currently being very clear with AIDE is essential for it to understand the context; but developing and communicating  a clear understanding of the problem is one of the essentially complex problems in software engineering. 

\section{What's Next}
The AIDE revolution has just begun, leaving open questions on what to expect in future. 

{\bf Language Models, Data, and Computational Power} The rapid progress in the capabilities of language models is difficult to quantify. A simple proxy for it is the increasing number of parameters in the language models presented by OpenAI in the last few years. In 2018 GPT-1 had 117M parameters. One year later GPT-2 pushed the boundaries to 1.5B, and in 2020 GPT-3 reached 175B parameters. Rumors place the next release (GPT-4) at an astonishing 100T parameters (500 $\times$ GPT-3) \citep{gpt4}. Similarly, the amount of training data available for code-related tasks is increasing every day, as are the computational capabilities of GPUs. Put together, these advances are expected to substantially improve the support AIDEs can provide to developers. To what extent will these improvements be affordable (in money and climate terms), and accessible (for those with no data centres)?

{\bf Improving the Quality of Training Data} As previously discussed, one of the main challenges when dealing with data-driven assistants lies in the quality of the training data. Manually checking all training instances is just not an option, but can AI help AI? In other words, can we teach AI what a high-quality training instance is? Whatever the underlying technology will be, defining techniques to automatically filter out noisy and flawed training instances is a cornerstone for AIDEs, and a focus for GitHub's next iteration of Copilot.

{\bf Code is Not (Just) Text} Language models have been proposed in the context of natural language processing, in which they are fed with a stream of tokens representing the text to process. However, code is not just text and there is active research investigating what the best representation is when feeding code as input to language models. For example, structural information can be extracted from the code Abstract Syntax Tree (AST) and used to boost the model's performance. 

{\bf Consumer-related Customization} While Copilot is able to provide code recommendations that are tailored for the specific coding task at hand, no customization is performed when it comes to the developer receiving such recommendation. However, two developers having a different technical background, coding history, and skills, may benefit from different recommendations. For example, more expert developers working on real-time software are likely to appreciate multi-threading solutions to a given task, while newcomers may be confused by its usage. Customizing the recommendations based on the target developer can substantially increase the usefulness of AIDEs.

{\bf AIDE Learning Rate} A last point worth discussing is the learning rate we can expect from AIDEs, namely the pace at which they'll be able to improve their capabilities. All the above-discussed points can contribute to that, but it's unclear when the AI will be able to pass a \emph{programmer's Turing Test}, for example submitting pull requests that reviewers cannot distinguish from a human's submission.

An important truism in software development is Fred Brook's maxim ``there is no silver bullet", derived from his insight into essential (inherent) complexity versus accidental (self-imposed) complexity. Like any new approach to our challenging field, AIDE is unlikely to become a panacea for software development. But it does seem to portend an important shift in how we develop software, and just might remove some of the accidental complexity in our projects.

\bibliographystyle{abbrvnat}
\bibliography{copilot.bib}

\begin{IEEEbiography}
{Neil A Ernst}{\,}is with University of Victoria, Canada. He conducts research into software design and natural language understanding in software. Contact him at \url{nernst@uvic.ca}.\\
\includegraphics[width=100px]{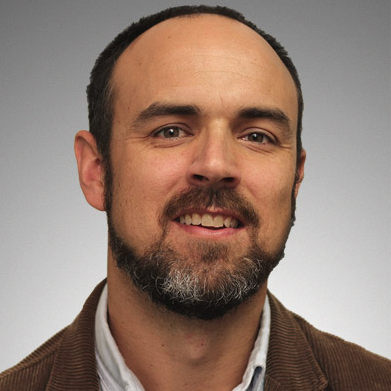}

\end{IEEEbiography}

\begin{IEEEbiography}
{Gabriele Bavota}{\,}is with Universit\`a della Svizzera italiana, Switzerland. He conducts research into mining software repositories and recommender systems for software developers.  Contact him at \url{gbavota@usi.ch}.\\
\includegraphics[width=100px]{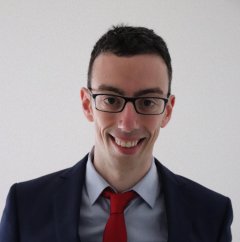}

\end{IEEEbiography}

\end{document}